\documentclass[12pt,letterpaper]{article}
\usepackage{amsmath,amssymb,pgf,pgfarrows,pgfnodes,float,appendix, hyperref}
\usepackage{graphicx}
\usepackage{subfigure}
\usepackage[margin=0.9in]{geometry}

\newcommand{\be}{\begin{equation}}
\newcommand{\ee}{\end{equation}}
\newcommand{\bea}{\begin{eqnarray}}
\newcommand{\eea}{\end{eqnarray}}

\title{{\rm\footnotesize \qquad \qquad \qquad \qquad \qquad \ \qquad \qquad \qquad \ \ \ \ \ \                  RUNHETC-2022-33}\vskip.5in   Old Ideas for New Physicists:1}
\author{Tom Banks\\
Department of Physics and NHETC\\
Rutgers University, Piscataway, NJ 08854\\
E-mail: \href{mailto:tibanks@ucsc.edu}{tibanks@ucsc.edu}
\\
\\
}
\date{}
\begin{document}
\maketitle

\begin{abstract} We review and clarify ideas proposed many years ago for understanding cosmology in a holographic framework.  The basic strategy is to use Jacobson's\cite{ted95} identification of Einstein's equations with the hydrodynamic equations of the "Area = 4 Entropy" law for causal diamonds, to identify a quantum system whose hydrodynamics match those of a given space-time.  This can be done exactly for a system with any positive cosmological constant, which saturates the entropy bound for all times. The quantum system is a sequence of (cut-off)  $1+1$ dimensional CFTs, with central charge proportional to the entropy in causal diamonds along an FRW geodesic.  This matches with a recent\cite{BZ} proposal that the modular Hamiltonian of any causal diamond for non-negative c.c. is the $L_0$ generator of such a CFT.  When an early de Sitter era is followed by slow roll expansion of the horizon, disjoint horizon volumes (which are gauge copies in an eternal dS space, but become physical due to slow roll expansion of the horizon area) manifest as a dilute gas of black holes in a post-inflationary era.  Entropy fluctuations of individual black holes manifest as CMB fluctuations, and the tensor/scalar ratio is suppressed by an extra factor of the slow roll parameter $\epsilon$. Black hole evaporation provides the Hot Big Bang and baryogenesis.  Black hole mergers can easily provide a source of primordial BH dark matter that dominates radiation at a temperature of $1$ eV, but numerical simulations are required to determine whether the model can explain the actual dark matter in our universe.

\end{abstract}
\maketitle

\section{Introduction}

Over two decades ago, W. Fischler and the author began construction of a general local theory of quantum gravity\cite{tbwf} based on the principle that a causal diamond in space time, with finite area holographic screen, was a finite dimensional subsystem of a quantum gravity model describing any larger region, with its own local dynamics, and entropy given by one quarter of the area of the screen in Planck units.  In the original work the density matrix of any diamond was assumed to be maximally uncertain.  While this extreme idea has been challenged by recent developments, it led to one of the most important conclusions of this body of work\cite{bfm}, namely that energy density localized near the geodesic in the diamond, must correspond to constraints on the state of the system: a typical\footnote{Throughout this note we will be using typical in the sense of the Eigenstate Thermalization Hypothesis\cite{ETH} rather than in the sense of Haar randomness.} state looks like empty space.  

The purpose of these two notes is to summarize these ideas for younger readers who were not around for their original dissemination, and to update them and eliminate misconsceptions that arose in many of our early papers.  They will be based entirely on semi-classical considerations, for the most part without use of the detailed formalism of H(olographic) S(pace) T(ime)\cite{HST}. The first note will concentrate on general principles and the application of those principles to a theory of inflationary cosmology that avoids the inconsistencies\footnote{These inconsistencies are often described by the phrase "transplanckian problems".  We view the fundamental problem as the violation of a conjectured bound\cite{CKN} on the number of states of field theory in a causal diamond, that are consistent approximations to states in a theory of quantum gravity.  
Field theory models of inflation assume there is no such bound.  The HST model evades the bound by using the fact that a theory of inflation is NOT even approximately, the same as a theory of de Sitter space.  In a formalism obeying the Covariant Entropy Bound, space-like separated causal diamonds in dS space whose past traces back to a single diamond of finite area, are not independent quantum systems.   However, the slow roll metric of inflationary cosmology expands the horizon, so that "causally separated diamonds of an approximately dS past", can be considered as independent quantum systems.}  of the field theoretic approach to inflation.  This theory gives predictions at odds with those of the field theoretic approach, but, as a consequence of our ignorance of the slow roll metric that is part of {\it any} modern theory of inflation, the two formalisms can not yet be distinguished by observation.  

The second note will be devoted to the idea that supersymmetry (SUSY) breaking in the real world is a consequence of interactions with the cosmological horizon of our asymptotically de Sitter (dS) space.  This has multiple implications for the phenomenology of particle physics, though in its current state the models based on this idea suffer from the {\it little hierarchy problem}.

I will assume the reader is familiar with General Relativity.   A causal diamond in a Lorentzian space-time is the intersection of the backward lightcone from some point on a time-like trajectory, and the future lightcone of a trajectory point of smaller proper time. Conversely, one can think of a time-like trajectory (in multiple ways) as a nested sequence of diamonds, corresponding to a nested sequence of proper time intervals. Time-like trajectories are important because they are the trajectories of idealized "detectors", the repository of all possible experimental information to which a physical model could be compared.   The dual constraints of quantum mechanics and general relativity assure us that no such idealized detector can exist.  We must increase its mass to minimize quantum fluctuations of its trajectory\footnote{In models of quantum gravity, particle trajectories are much more well defined than in ordinary quantum mechanics of quantum field theory, because they are properties of large subsystems (jets), consisting of a particle and soft gravitons it emits, or absorbs, whose asymptotic momenta lie within the opening angle of a cone.}, but we cannot do this without distorting the environment which it is supposed to measure.   If we insist on a fixed degree of localization of detector energy density, then we will form a black hole if the detector mass exceeds the Planck mass. Black holes have only limited detection capacity because of the no hair theorem.  Nonetheless we will take the mathematical abstraction of a time-like trajectory as an acceptable idealization of reality.  Refusal to do so in models of quantum gravity restricts attention to asymptotic observables in space-times that do not resemble the one in which we appear to live.

Any causal diamond has two geometric invariants associated with it.  The first is the maximal proper time between its time-like separated tips.  The second is the maximal area of a $d-2$ surface in a null foliation of the diamond boundary.  The diamond boundary is not a differentiable manifold, but a gluing together of two null manifolds with metrics \begin{equation} ds_{\pm}^2 = g_{ij} (u_{\pm}, y) dy^i dy^j  + du_{\pm} dy^i A_i (u_{\pm}, y) . \end{equation}   The maximal $d - 2$ volume, henceforth {\it area} $A$, of the $d - 2$ dimensional Riemannian metrics on the boundary, is called {\it the area of the holographic screen, or holoscreen, of the diamond}.  In many cases, the holoscreen is the bifurcation surface between the past and future null boundaries of the diamond.

For small enough diamonds, the relation between area and proper time is of the form $A \sim t^{d-2}$, while for large proper times the relation depends crucially on the cosmological constant (c.c.).  For positive c.c. $A$ remains finite while $t $ goes to infinity, while for negative c.c.  $A$ becomes infinite for a time of order the AdS radius.  
For vanishing c.c., the power law relation persists to infinite time.  These differences mirror another feature of gravitational physics: the black hole spectrum in space-times with positive c.c. is bounded, with a bound determined by the c.c..  For negative c.c. the logarithm of the spectral density grows like a power of the energy smaller than $1$, with a coefficient that depends on the c.c. , while for vanishing c.c. the power is greater than one.  The connection between these two observations is a consequence of the relation between the size and energy of a black hole.  
Both of them lead to the conclusion, abundantly verified by the AdS/CFT correspondence, that different values of the c.c. correspond to different models of quantum gravity and that the c.c. is unaffected by field theoretic renormalization.  It is not a running EFT parameter, but rather an asymptotic (in large energy or distance) boundary condition defining a model.

\section{Causal Diamonds as Quantum Subsystems}

In quantum field theory, causal diamonds correspond to subalgebras of the full operator algebra, which are Type III Murray-von Neumann factors.  This means that the only operators in the center of the algebra are multiples of the identity, and that the diamond algebra has a non-trivial commutant in the full algebra.  Furthermore it is the commutant of its commutant.  Type III implies that every projection in the algebra has trace $0$ or $\infty$ and that there are no pure states, or even density matrices, on the algebra.  The latter fact can be easily understood.  Commutators of local operators are singular on the light cone, so the operators inside the diamond algebra but localized near the boundary are infinitely entangled with those an infinitesimal time-like interval from them, outside the diamond.
The entanglement entropy between the diamond interior and the rest of the operator algebra is infinite, proportional to the area of the diamond multiplied by a power of the UV cutoff\cite{sred}\cite{cw}\cite{sorkin}.  Despite the fact that it is power law in the cutoff, the authors of\cite{CHM} showed that it could be computed in a universal way.   It is independent of the quantum state of the system, as long as that state is created by a finite number of smeared local operators, with smearing functions having support space-like separated from the diamond boundary.  The coefficient of the divergence depends on the details of the quantum field theory\cite{CHM}.

Heisenberg evolution, on "diamond universe" DU time slices\cite{CHM}\cite{JV} inside the diamond, maps the diamond subalgebra into itself.  As a consequence of the properties of Type III algebras, it is an automorphism of the algebra, but is not implemented by a unitary transformation belonging to the algebra.   

Consider a quantum field theory that can be viewed as a limit of lattice models where each site Hilbert space is finite dimensional, and the range of interactions is finite.  Any such model has a maximal Lieb-Robinson velocity\cite{liebrobinson}\footnote{Lieb and Robinson showed that in lattice systems with finite dimensional site Hilbert spaces and finite range couplings, the operator norm of the commutator of two Heisenberg operators on different sites,$x,y$ was bounded by $e^{- a [d(x,y) - v_{LR} t]}$ for large distance, $d(x,y)$ between the points.  The "constant" $a$ is actually an increasing function of distance.}, and we can define (up to exponentially small corrections) finite dimensional subalgebras of the full operator algebra, corresponding to LR "causal diamonds".  Heisenberg evolution using the full Hamiltonian of the model takes these algebras into themselves, over periods of time that are shorter than the distance of an operator to the boundary divided by the maximal LR velocity.  This is the analog of the field theory results in DU coordinates, but since the diamond algebras are finite dimensional the Heisenberg automorphism is unitarily implementable.  That is, there is a time dependent Hamiltonian, constructed from operators in the diamond algebra, which mimics the full Heisenberg evolution whenever the operator remains in the diamond.  

Starting from a lattice point, we can build up a sequence of Hilbert spaces, which describe the nested diamonds generated by operators at that point, over a sequence of nested "proper time" intervals.  Heisenberg evolution will be implemented by a proper time dependent Hamiltonian, which preserves "LR causality"  in this sequence of spaces.    Now imagine starting at another lattice point and following the same procedure.  At some time, the LR diamonds of the two points will overlap.   The overlap defines a subset of q-bit operators, which belong to both diamond subalgebras.   The fact that everything was generated by a single global Hamiltonian evolution guarantees that expectation values of products of those overlap operators will be the same, independently of which point we start from.  

Thus, one can formulate lattice field theory dynamics in terms of multiple evolutions in nested LR causal diamonds.  The overlap conditions are the statement that the maximal area diamond in the overlap between the two diamonds is identified with a subsystem in each, and the density matrices on these subsystems should have the same entanglement spectra, whichever sequence of nested diamonds one uses to compute them.  This infinite set of intricate conditions, is automatically guaranteed by the existence of the global field theory Hamiltonian, which generates the Heisenberg automorphisms in each nested sequence of LR diamonds simultaneously.

Our basic claim, for almost two decades, has been that the facts above about the formulation of dynamics in nested causal diamonds, do not contradict anything we know about quantum gravity.  Perturbative string theory and AdS/CFT describe only correlation functions of operators on infinite boundaries of space-time with frozen asymptotic geometries.  They can be related to local physics only in perturbation theory around classical geometries.  The tensor network\cite{tn} formulation of AdS/CFT allows for localization on scales larger than the AdS radius only.  The reason for this is that local physics intrinsically requires a choice of gauge.

Our current understanding of the principle of general covariance is a mathematization of Einstein's original idea that detectors following different world lines should have equivalent descriptions of physical processes that were accessible to both detectors.  This principle, and the principle that a detector could access only a finite amount of information in a finite proper time, are the basis for both the Special and General Theories of Relativity.   {\it It follows that a local view of physics corresponds to a choice of gauge.   The natural local regions are causal diamonds.}   We should think of covering space-time with a particular set of nested diamonds as a particular choice of physical gauge and the equality of entanglement spectra for the density matrices of overlap diamonds as the {\it Quantum Principle of General Relativity}.

The place where quantum gravity differs from QFT is in the number of q-bits associated with a given diamond, and their localization inside the diamond.  There is also not, as yet, any clear indication of whether the sort of global Hamiltonian description of the system that we have in QFT is available.  The QFT calculations of entanglement entropy of a diamond put a lower bound on the dimension of the Hilbert space of the diamond.  They also suggest that the QFT description of the states with maximal weight in the density matrix is flawed.  Indeed, those calculations were motivated by conjectures\cite{ct}\cite{su} that they should be thought of as renormalization of Newton's constant in the Bekenstein-Hawking formula for the entropy of a black hole.  This means that the actual explanation for the entanglement entropy is a Planck scale problem, not addressable by QFT.  We'll see more evidence for that when we discuss entropy fluctuations below.

An upper bound on the amount of information in a diamond describable by QFT was discussed long ago\cite{CKN}.  The argument is phrased in terms of crude UV and IR cutoffs in a fixed low curvature diamond, but its conclusion can be stated in a covariant form and similar results are obtained for more sophisticated cutoffs\cite{bd}\cite{detal}.  
In field theory, the entropy and energy in a region of size $R$  in $d$ dimensions, scale like
\begin{equation} S \sim (M_c R)^{d-1}\ \ \ \ \ E \sim M_c (M_c R)^{d-1} . \end{equation}
Insisting that we should ignore field theory states where $R$ is smaller than the Schwarzschild radius corresponding to the energy $E$ we get 
\begin{equation} S < (A/4G_N)^{\frac{d-1}{d}} \ll A/4 G_N . \end{equation}
That is, field theory entropy can never account for the Bekenstein Hawking entropy.  We can phrase this as a covariant bound
\begin{equation} S_{QFT} < K S_{BH}^{\frac{d-1}{d}} , \end{equation} on the QFT entropy in a diamond.  It's unclear how to calculate the constant $K$.  Bounds of this form were noted by previous authors, particularly 't Hooft.  A fascinating observation of of\cite{CKN} was that one could impose the requisite restrictions on QFT states without affecting any of the precise agreement of QFT with experiment.  The authors of\cite{bd}\cite{detal} showed that the requisite depletion of degrees of freedom can be done in a way that yields only $G_N$-suppressed contributions to observables in perturbative QFT\footnote{In light front gauge perturbative string theory, an infrared cutoff implies quantization of longitudinal momentum, breaking the string into bits. States with with low light front energy have large longitudinal momentum and fixed transverse momentum. It's clear that, although low energy few particle S-matrix elements are well approximated by effective field theory, the nature of the space of physical states is drastically different.  It is not clear whether any sort of semi-perturbative analysis could lead to the kind of restrictions envisaged in\cite{bd}\cite{detal}.}

In field theory, the entropy appears to come from field modes of very short wavelength, propagating near the boundary of the diamond, but this is an illusion.  In the global field theory vacuum state, those modes are locked into maximal entanglement with modes just outside the boundary.  Any attempt to disentangle them and excite them into a random state, costs enormous energy as measured by the field theory Hamiltonian.  The basis of quantum statistical mechanics is the assertion that many systems have dense bands of energy levels that share the same macroscopic features, so that the quantum expectation values of simple operators are the same as ensemble averages.   While the entanglement entropy of a diamond in a specific QFT is universal, that universality does not have anything to do with excitations inside the diamond.  In QFT, if we take generic states inside the diamond the entanglement entropy no longer has an area law.  The CKN argument tells us that once one takes gravity into account, most field theoretic states inside the diamond will form black holes, and the transition occurs when the field theory entropy is much less than the black hole entropy.  

These remarks motivate the Covariant Entropy Principle\cite{fsb} : {\it In models of Quantum Gravity the entropy in a diamond is given by the Bekenstein Hawking formula, and consists of states that have small energy density outside a small region near the diamond boundary.   These states are very densely spaced and all describe the same macroscopic properties of space-time.}  In the original version of this conjecture\cite{tbwf} the density matrix of these states was taken to be maximally uncertain.  The strongest argument for this came (anachronistically) from Jacobson's\cite{ted95} derivation of Einstein's equations from this principle, as the hydrodynamics of space-time.  Jacobson studies the local first law of thermodynamics, from the point of view of a maximally accelerated observer whose trajectory reverses direction at a fixed point in space-time.  Interpreting maximal acceleration as infinite temperature, one is led to the conclusion that the density matrix is maximally uncertain.  In a theory of QG however, one can imagine that maximal temperature means something of order Planck scale, which leaves room for entropy fluctuations\footnote{In fact, I believe the infinite temperature argument is a red herring.  What really matters is the spectrum of the dimensionless modular Hamiltonian, $K$, of the diamond. Different trajectories $i$, write this Hamiltonian as $K = \beta_i H_i$ and see redshifts/blueshifts of energy and temperature, because of time dilaton.}.  We'll argue below that there is indeed evidence that diamond boundaries have universal entropy fluctuations. 

\subsection{Euclidean Path Integrals} 

In a famous paper, Gibbons and Hawking\cite{gh1} calculated black hole entropy in terms of Euclidean path integrals.  This can be extended to de Sitter space\cite{gh2} and the entropic interpretation of Euclidean action also illuminates the study of Coleman Deluccia\cite{cdl}\cite{tbheretic}\cite{bw} instantons describing transitions between two dS spaces.  Many other Euclidean calculations of entropy, including that of empty causal diamonds, have appeared since then\cite{bdf}\cite{df}\cite{maldleuk}.  

From the traditional string theory point of view, where Einstein's action is viewed as a low energy EFT, it's somewhat surprising that such classical calculations know something about entropy, which is a property of the spectrum of a quantum Hamiltonian in a regime far from the ground state.  Jacobson's\cite{ted95} interpretation of Einstein's equations as the hydrodynamics of quantum gravity sheds light on this mystery.  Hydrodynamics is derived from QM\cite{bl}\cite{hong}\cite{ranga} by finding large sets of commuting, approximately conserved operators.  In systems well modeled by lattice field theory, these are the integrals of the time components of conserved currents, over blocks of the lattice much larger than microscopic scales.  The expectation values of these special operators satisfy stochastic hydrodynamic equations up to corrections that are powers of the ratio of microscopic to macroscopic length scales.   The time scales over which hydrodynamic operators vary is much longer than microscopic time scales, but much shorter than classical recurrence times of the subsystem over which they're averaged.  Thus, in the hydrodynamic equations a local entropy density appears via the average of various matrix elements over energy bins larger than the inverse recurrence time but smaller than the inverse hydrodynamic time scale.  If GR is viewed as hydrodynamics, it is not surprising that it contains a formula for entropy.  Note that there is no contradiction between this interpretation of GR and its use as an effective field theory to describe low energy long wavelength fluctuations in a system with a ground state: phonons and other hydrodynamic quasiparticles dominate the low energy physics of most condensed matter systems.

\section{Where's the Bits???}

\begin{figure}[H]
\centering
\includegraphics[scale=0.3]{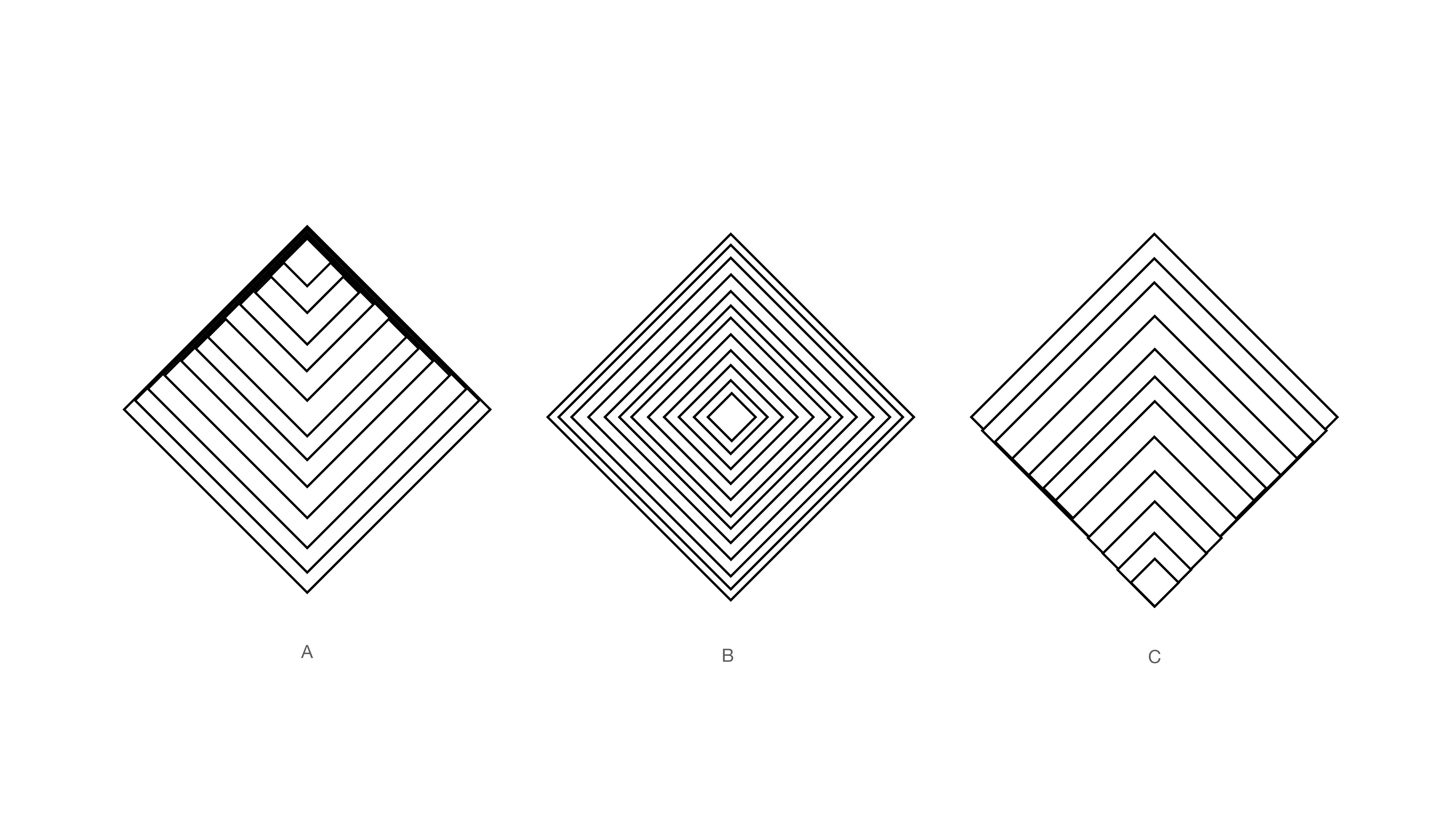}
\caption{Past/Future Directed and Time Symmetric Causal Diamonds}
\label{nests}
\end{figure}

To summarize the previous section in a sentence: an empty causal diamond in a model of quantum gravity has a finite dimensional Hilbert space if the area of its holoscreen is finite, and most of the states in that Hilbert space are localized near the boundary diamond and therefore have very low energy in the vicinity of the diamond's geodesic.
This however is not the whole story.  Cover the inside of the diamond with a nested set of diamonds whose time-like tips are separated by a single Planck time.  This nest of diamonds can be past oriented, future oriented or time symmetric\ref{nests}.   

Our axioms insist that each diamond in the nest is a subsystem of the next, and that the proper time dependent Hamiltonian along the trajectory defined by the nest entangles that subsystem with the new degrees of freedom that are added in the larger diamond.  For diamonds much smaller than the radius of curvature defined by the c.c., which we will call {\it small causal diamonds}, the number of DOF that's added to a diamond of radius $R$ in a Planck time is of order $(R/L_P)^{d-3}$.

In order to understand states localized within the bulk of, rather than on the boundary of, diamonds, we need two clues from semi-classical black hole physics.  The first is the entropy formula for Schwarzschild de Sitter (dS) black holes.  In $4$ dimensions, the metric is
\begin{equation} ds^2 = - f(r) dt^2 + \frac{dr^2}{f(r)} + r^2 d\Omega_{2}^2 , \end{equation}
\begin{equation}  f(r) = (1 - \frac{2ML_P^2}{r} - \frac{r^2}{R^2}) . \end{equation}   The blackening factor $f(r)$ vanishes at the roots of the cubic polynomial
\begin{equation} 0 = (r - R_+) (r - R_-) (r + R_+ + R_-) , \end{equation} where
\begin{equation} R_+ R_- (R_+ + R_-)= 2M L_P^2 R^2, \end{equation} and
\begin{equation} R_+^2 + R_-^2 + R_+ R_- = R^2 . \end{equation}
When $R_- \ll R_+$ the black hole has an entropy deficit relative to empty dS space, according to the Bekenstein-Gibbons-Hawking formulae, of 
\begin{equation} \Delta S \approx - 2\pi M R . \end{equation}   This is a derivation of the Gibbons Hawking temperature of dS space without quantum field theory.  It tells us that the empty dS vacuum is the maximal entropy state of the system and that localized states are constrained.  Note that we've left the black hole entropy out of this formula, as would be appropriate for an elementary system with a mass much less than the Planck mass.  It would have the same Schwarzschild field, except very near the black hole horizon.

In the simplest interpretation one identifies the maximal entropy as the logarithm of the dimension $D_R$  of the Hilbert space, and the states with localized static objects as states constrained to a subspace of dimension approximately $D_M = e^{-2\pi M R} D_R$, with the Boltzmann factor interpreted as the probability that a random state has a projection of order $1$ on the constrained subspace.   Later we'll argue that the density matrix of empty dS space has entropy fluctuations $\Delta S = S^{1/2} $.

The second semi-classical fact that makes the same point is the black hole entropy formula for all {\it small} black holes.  Consider a particle of small mass $m$ somewhere in the vicinity of a black hole of mass $M \gg M_P$.  The entropy of the combined system, if the particle falls into the black hole,  is  $(\frac{M + m}{M_P})^{\frac{d-2}{d-3}} A_{d-2}/4$, ($A_{d-2}$ is the area of a unit $d-2$ sphere), much larger than the initial black hole entropy.  This is most easily understood if one assumes that the Hilbert space of the particle plus black hole system had $\frac{(d-2)}{(d -3} (\frac{M}{M_P})^{\frac{1}{d-3}} (m/M_P)$ frozen q-bits before the particle fell in, and the process of "falling in" consists of equilibrating those degrees of freedom to make a more generic state of the larger black hole.  Furthermore, if we assume that in states where those degrees of freedom were frozen, the particle and black hole subsystems had vanishing interaction\footnote{All of these statements are meant at leading order in $m/M_{BH}$.  There are interactions mediated by virtually unfreezing and freezing the constrained variables at second order in perturbation theory.}  then we can understand why there is a long period of time during which the particle is oblivious to the existence of the states on the black hole horizon.  The time scale for the beginning of equilibration is the number of operations of the Hamiltonian it takes to unfreeze all those degrees of freedom, times the inverse of the "typical" scale of eigenvalue differences.   One must again be careful about the word typical.  Large systems generally have a microscopic scale of energy differences, which determines their short time dynamics, and then longer "hydrodynamic" and "recurrence" scales.  Typical, in our usage, will mean the short time dynamical scale.   Because of the Unruh effect, the short time scale depends on the trajectory whose proper time we are using.  For the dynamics of most of the degrees of freedom in a diamond, that scale varies between the inverse diamond size $R^{-1}$ for the geodesic, and the Planck scale for a "maximally accelerated" trajectory. 

Thus black hole entropy formulae for both SdS and ordinary flat space black holes suggest that localized excitations inside a diamond are constrained, out of equilibrium states of the holographic variables on the diamond boundary.  To researchers used to the description of local excitations in the AdS/CFT correspondence, this prescription may seem strange.  To reconcile the two points of view we can turn to the tensor network description of AdS space.   This is usually described on a fixed time slice, with a fixed choice of conformal generator $K_0 + P_0$ in its conjugacy class chosen to be the Hamiltonian.  This determines a unique timelike geodesic in AdS, whose proper time is related to CFT time by an infinite rescaling.  If we cut off the infinite volume, the tensor network is a sequence of lattice theories.  The scale of the lattice spacing is much smaller than the AdS radius, and the number of q-bits in the site Hilbert space is the area of a sphere of AdS radius\cite{susslargenode}\footnote{If, as in all known examples, there are compact dimensions whose size is of order the AdS radius, then the sphere is higher dimensional.}.  The lattices live on concentric spheres of smaller and smaller radius, but their actual physical radius is not the number of lattice points in a diameter, but rather the number of AdS radii.   At the center of this contraption there is a single sphere, which describes the base of  causal diamonds along the central trajectory, over times for which the area of the diamond is less than the AdS area.  
The tensor network construction entangles the central sphere with the much larger system of lattices surrounding it.  Thus, by Page's theorem\cite{page}, for most states of the system, the central sphere subsystem will be in a density matrix of large entropy. 
Furthermore, that state will not be significantly affected by most perturbations on the boundary.   It is protected, because a tensor network is an error correcting code\cite{ecc}.   Thus, the state of a small diamond in AdS space\footnote{The isometries of AdS can map any small diamond into the central diamond in a tensor network description.} is like the vacuum of dS or a small finite diamond of any kind.   The special perturbations that can inject "particles" into this diamond are complex operators, which impose a large number of constraints on the diamond Hilbert space.
 
\subsection{Area Preserving Diffeomorphism Invariance and Fast Scrambling}

There is a geometric, or more properly a topological measure theoretic, way of understanding these relations.  Suppose that we can associate the operators that appear in the Hamiltonian with sub-regions on a topological $d - 2$ sphere of area $A_{d-2} (R/L_P)^{d-2}$, where $A_{d-2}$ is the area of the unit sphere.  A state with a localized object in it consists of a small area $A_{d-2} a^{d-2}$, $ a\ll R$ surrounded by a topological annulus of area  $(d-3) A_{d-2} a^{d-3} R$, in which all the operators appearing in the Hamiltonian vanish.  If the Hamiltonian is an integral of a product of operators at the same point, then that vanishing leads to a decoupling of the region $a$ from most of the system.  

This description of the interactions makes it look like the vanishing of a subset of variables defines a conserved subspace of states.   It also seems to involve an infinite number of degrees of freedom, since the variables are described as generalized functions on a topological space with the topology of a sphere.  Both of these problems are ameliorated by "fuzzifying" the space.   Take a complete set of square integrable functions $Y_L (\Omega) $ on the sphere, and write a field as
\begin{equation} f(\Omega) = \sum a_L Y_L (\Omega) , \end{equation}  with the $a_L$ taken to be fermionic oscillators.  Cut off the sum over $L$ so that the maximal fermion entropy is equal to the area of the sphere in Planck units.   The anti-commutators
\begin{equation} [f(\Omega), f(\Omega^{\prime})]_+ , \end{equation} are non-zero for almost all pairs of points.   If the $Y_L$ are eigensections of the Dirac operator (with the round metric) on the sphere, then the anti-commutator vanishes like $d^{- \frac{d-2}{2}}$ with $ d(\Omega, \Omega^{\prime})$ the distance between the two points in the round metric.  Although we cannot construct a characteristic function of an annulus for finite $L_{max}$ we can get close if $L_{max}$ is large.  So the measure theoretic intuition about decoupling is valid if $L_{max} \gg 1$, but the states where "functions vanish on an annulus" are not true eigenstates of the Hamiltonian.   The long range of the anti-commutator of "fuzzily local fields" , implies that a Hamiltonian built in terms of products of $f-local$ fields will not obey the ballistic limits on scrambling implied by the Lieb-Robinson bound.  

In the limit $L \rightarrow\infty$, it appears that the fields become local, but if the Hamiltonian has the form $$ \int_{S^{d-2}} P(M_i)$$ , where the $M_i$ are differential forms constructed as bilinears in the Dirac eigensections and the product in the polynomial $P$ is a wedge product, then the system is formally invariant under area preserving diffeomorphisms\footnote{It's important that these diffeomorphisms are global, not gauge, symmetries.} and cannot obey the constraints of locality for any specific metric on the sphere.  In four dimensions, the cutoff Dirac spectrum is described by fermions labelled by the $[N]\otimes[N + 1]$ dimensional representation $\psi_i^A$ of $SU(2)$, the bilinears are $M_i^j = \psi_i^A \psi^{\dagger\ j}_A$ and the usual fuzzy approximation to the group of area preserving diffeos emerges if we take the Hamiltonian to be a single trace of a polynomial in the matrices $M$.  A similar proposal has been made for higher dimensions\cite{tbwfhigherd}, but not all the kinks have been worked out.

\subsection{Tensor Networks and AdS/CFT}

As hinted above, in systems with AdS asymptotics, we have to modify the principles discussed above.  Ultimately, this has to do with the fact that the area of a diamond in asymptotically AdS spacetimes goes to infinity in finite proper time.  The conformal boundary is a timelike surface and we have to define a quantum system with an infinite dimensional Hilbert space to propagate along that boundary.  The system must be a representation of the conformal group of the boundary, and the results of\cite{gkpw} show us that if the theory in the bulk has dynamical gravity then the boundary theory will have a conserved local stress tensor, whose integrals define the conformal generators.  In short, the boundary theory is a CFT.

Maldacena\cite{scaleradius} and the authors of\cite{witsuss}, argued that regions enclosed by diamonds of finite area in the bulk, were some sort of cutoff version of the CFT.  Swingle\cite{tn} first proposed that the cutoff be a tensor network, a sequence of lattice models that converge on the CFT via "entanglement renormalization"\cite{tnrg} .  The tensor network formulation of CFTs is essentially equivalent to the statement that the boundary theory is an error correcting quantum code for information about the bulk\cite{ecc}. Fischler and the present author proposed that the entangling transformations be time evolution operators in causal diamonds along a particular time-like geodesic in AdS space\cite{tbwfads}, picked out by the choice of Hamiltonian among the generators of the conformal group.

Susskind\cite{susslargenode} argued that for an AdS radius much larger than microscopic scales\footnote{It has been known since the work of 't Hooft that certain large N gauge theories are dual to string theories, with the $1/N$ expansion implying that the string length is much larger than the Planck length.  Obtaining model CFTs where the AdS radius is much larger than the string scale, so that the bulk Einstein-Hilbert Lagrangian is a good approximation to physical correlators, is much more difficult.  All examples appear to require supersymmetry and at least two compact dimensions with radii of order the AdS radius.} one always needed the nodes of the tensor network to be large Hilbert spaces.  This can be stated in terms of the CEP:  to describe causal diamonds of area $R^{D-2}$, where $D$ is the number of dimensions whose size is of order the AdS radius, we need at least $(R/L_P)^{d-2}$ q-bits.  The proposal\cite{tbwfads}\cite{BZ} is that the dynamics inside each of the tensor nodes is that of the {\it small diamond} models described above, while different nodes are coupled to each other locally on a lattice.   Using the connection between thermalization time and infall time to the singularity outlined above, we conclude that infall time should be of order the AdS radius, and independent of the size of the black hole, for large black holes in AdS.   We also conclude that the spectrum of quasi-normal modes for a large AdS black hole should include long lived sound waves, unlike {\it small} black holes, for any value of the c.c. . Of course, the last conclusion is implied by the AdS/CFT duality between large black holes and thermal states of the boundary CFT, and the first is a well known geometric property of large AdS black holes.  

\subsection{Fast Scrambling}

If a particle is dropped into a black hole, its charge and mass are recorded in the electric and gravitational fields on the {\it stretched horizon}\cite{membrane} .  The stretched horizon is a timelike surface that is of order Planck distance from the event horizon. The originally localized charge/mass densities decay exponentially, because the stretched horizon is accelerated.   For small spherically symmetric black holes the charge distribution becomes spherically symmetric in a time of order $R_S {\rm ln}\ (R_S/L_P)$ where the Planck length enters because we consider inhomogeneous fields the size of typical quantum fluctuations to be negligible.  While this observation was at least implicit in the original Membrane Paradigm literature, its significance for black hole quantum mechanics was first emphasized in\cite{susslind}.  Hayden and Preskill\cite{hp} argued that if information was not scrambled on time scales of order $R_S  {\rm ln}\ (R_S/L_P)$ , then the resolution of the black hole information "paradox" by Black Hole Complementarity\cite{su} would be untenable.  Susskind and Sekino\cite{ss} then showed that a variety of quantum systems could be fast scramblers.  A typical example is a collection of q-bits with k-local Hamiltonians, which couple at most $k$ q-bits together.  If there are no locality restrictions on the couplings, then the system is a fast scrambler.  Each action of the Hamiltonian in the Heisenberg equations, "infects" $k$ other q-bits, and so information spreads around the system exponentially fast.  It is likely that Hamiltonians that are single traces of matrix polynomials, with matrix elements that are bosonic or fermionic creation/annihilation operators are also fast scramblers.   Intuitively, this should follow from the invariance of these Hamiltonians under "fuzzy" volume preserving diffeomorphisms. In such a system a small spherical cap is equivalent to a thin "amoeba" of the same area, which can touch an arbitrary point on the sphere.  Scrambling of information cannot be limited by the distance between points on the sphere in any metric.   
 Lattice field theories are not fast scramblers, as proven by Lieb and Robinson\cite{liebrobinson}, but field theories on the sphere with sharp angular momentum cutoffs might well be.  The commutators of "local" operators fall off like power laws in the distance.  

\subsection{Executive Summary}

Analysis of semi-classical  entropy formulae, and quasi-normal modes indicates a universal local behavior of null surfaces in models of quantum gravity.  Semi-classical arguments\cite{CKN} and QFT calculations of the entanglement entropy of causal diamonds\cite{sred}\cite{cw}\cite{sorkin} suggest that the Bekenstein-Hawking formula for black hole entropy applies to the entropy of any causal diamond, identifying a finite area diamond as a finite dimensional subsystem in any model of QG.  The causal structure of QG is then identical to that of QFT, with only the relation between the total entropy and geometry distinguishing them.  This leads to a formulation of QG in terms of "many-fingered" proper time evolution along different time-like trajectories.  Causality is implemented by proper time dependent Hamiltonians, which couple together only the DOF in the causal diamond of a given time interval.   The principle of relativity is implemented by the requirement that the density matrices assigned to the maximal diamond in the intersection of diamonds along two different trajectories, have identical entanglement spectra.

The above principles localize quantum information on the boundaries of causal diamonds, with Hamiltonian evolution entangling the q-bits in one diamond with the new ones that are added every Planck time.  They do not tell us how to describe excitations localized in the bulk of a diamond.  Black hole entropy formulae suggest that such localized states are constrained states of the boundary system, with a number of constraints scaling like $ER$ where $R$ is the radius of the diamond and $E$ the localized energy.  We will discuss this further in the cosmological section below.

\section{Entropy Fluctuations}

As noted above, in early work\cite{tbwf} the entropy of a causal diamond was identified with that of the maximally uncertain density matrix on the Hilbert space of a diamond.  Quite frankly, the original argument was "What else could it be?", which was later superseded by an appeal to Jacobson's use\cite{ted95} of infinite temperature Unruh observers to derive Einstein's equations from the first law of thermodynamics and the BH entropy formula.  In fact however there was work from the 1990s which suggested an elegant, universal but not maximally uncertain prescription for the density matrix of a causal diamond.\cite{carlip}\cite{solodukhin}.  

Let me recapitulate this argument briefly, using the approach of Solodukhin\cite{solodukhin} as generalized to arbitrary causal diamond boundaries by\cite{BZ}.   We examine a causal diamond boundary in a general solution of the Einstein-Hilbert Lagrangian, which has the near horizon form
\begin{equation} ds^2 = (\eta_{ab} + h_{ab}) \times dy^i dy^J g^{(d-2)}_{ij} (y) . \end{equation}   We consider linearized fluctuations only away from the two dimensional Rindler metric.  This can be motivated by the logic in\cite{V291}: we are considering short wavelength fluctuations concentrated near the horizon, while the natural length scale of the transverse metric is macroscopic.   The dimensionally reduced action for fluctuations has terms proportional to the transverse volume, and to the integrated scalar curvature of the transverse metric.  We'll use the approach of Solodukhin\cite{solodukhin}, who observed that the linearized near horizon action had the form of the classical Liouville CFT, with classical central charge proportional to the transverse volume.  All other terms in the action are irrelevant perturbations of this CFT: they scale away as one approaches the horizon.

Rather than quantizing this two dimensional CFT directly, Carlip and Solodukhin borrowed the logic of Strominger's\cite{andy} {\it post facto} "derivation" of the $AdS_3/CFT_2$ correspondence from the Brown-Henneaux\cite{brownhenneaux} boundary Liouville action for $3$ dimensional EH gravity.  That is, they assumed the Virasoro algebra in the effective action was realized in the quantum theory in a representation where the $L_0$ generator is bounded from below.  Cardy's theorem then allows one to calculate the entropy, and it turns out to match the Bekenstein-Hawking formula for every black hole known to man.

The new observations of\cite{BZ} were that this derivation applied to arbitrary diamond boundaries with the above geometry, not just black holes, and that the assumption of $1 + 1$ dimensional CFT behavior implied that entropy fluctuations satisfied
\begin{equation} \langle (K - \langle K\rangle)^2 \rangle = \langle K \rangle . \end{equation}  $K$ is the modular Hamiltonian of the causal diamond. This is a universal rule and had been established for Ryu-Takayanagi (boundary anchored) diamonds in AdS space by Verlinde and Zurek\cite{VZ2}.  For RT diamonds the result can also be derived as a special case of Perlmutter's general formula\cite{perlprivate} for $(\Delta K)^2$ for CFTs, in the case where the CFT has an EH dual\cite{perlprivate}.  The authors of\cite{BZ} also showed that the formula for diamonds in Minkowski space followed from the RS2\cite{RS2} model of flat space coupled to gravity as a brane in AdS space.

The conjecture of\cite{BZ} is that these considerations apply, locally, to all diamond boundaries with the near horizon Rindler structure above.   For large diamonds in AdS one uses the tensor network description of the CFT.   The universal fluctuation formula applies to a single node of the tensor network.  A large $AdS_d$ black hole corresponds to a completely equilibrated state of some shell in the tensor network, where the adjective {\it large} implies a shell with many nodes.  Entropy fluctuations in this equilibrium state are dominated by states that are uniform over the shell and satisfy the universal formula for CFTs on $R \times S^{d-2}$\footnote{The TN description violates exact $SO(d-1)$ invariance of the classical black hole dynamics.  I hope to return to a resolution of this discrepancy in a future publication.}.   

A $1 + 1$ dimensional CFT does not of course have finite entropy.  In the work of Carlip and Solodukhin, the finiteness of the entropy comes from identifying the black hole entropy with the Cardy formula at a specific $L_0$ level, which is calculated from the classical solution of the Liouville action, with fixed transverse area.  The Cardy formula does not refer to exact degeneracies for a generic CFT, but rather is the smoothed density of states for a chaotic spectrum.  A picturesque way of thinking about this is to consider the CFT on a finite interval of length $I$, with a UV cutoff $M_c$ satisfying $M_c I = K_c \gg 1$ .   $K_c$ is independent of the central charge.   This picture was first proposed in\cite{holocosm1+1} and the interval was associated with a small longitudinal region in light front coordinates near a point on the horizon. At the moment there is no compelling evidence for or against this intuitive description, but we will adopt it for convenience.

The fact that the area of the causal diamond is encoded in the central charge of the CFT suggests that we should think of the geometry of the transverse surface is associated with the {\it target space} of the CFT.   This sort of encoding is the way in which the geometry of compact dimensions is described in AdS/CFT, as well as the way transverse dimensions appear in Matrix Theory\cite{bfss}.  Thinking first about free CFTs, we can think about the free fields as the expansion coefficients of a section of some vector bundle over the transverse manifold.  This will give a finite value of the central charge if we cut off the expansion, which can be done in a covariant way if the cutoff is imposed on the eigenvalues of some covariant differential operator like the Laplace operator.  In the holographic space-time (HST)\cite{HST} formalism we have chosen the spinor bundle, with a cutoff on the transverse Dirac operator.  This has many virtues, most of which are better explained in the second of these notes.  For the moment, let us just comment that we must have spinor operators if we want to have half integer spin particles in our model, and that bilinears in spinors are differential forms.

Susskind and Sekino\cite{ss} emphasized that semi-classical properties of black hole horizons, as well as the consistency of Black Hole Complementarity, required the quantum system representing the horizon to be a fast scrambler. Fischler and the present author suggested that fast scrambling was easily achieved if the Hamiltonian was invariant under area preserving diffeomorphisms of the horizon.  In such a system, a small localized area on the transverse manifold is equivalent to an equal area spread out in a very thin "amoeba" over the whole manifold.   Such a system cannot have a Lieb-Robinson bound on the propagation of information.  So a candidate Hamiltonian would be a sum of terms of the form
\begin{equation} \int \prod_p F_p \delta_{d-2,\sum p} , \end{equation} where the $F_p$ are products of bilinears in spinors, transforming as $p-$forms.  Quantizing the expansion coefficients of the spinors via\footnote{On transverse manifolds with non-trivial cycles, one can replace these with more general finite dimensional super-algebra anti-commutation relations.  We'll discuss this some more in the second of these notes.}
\begin{equation} [\psi_m^{\dagger} , \psi_n]_+ = \delta_{mn} . \end{equation}  Here $m,n$ label different eigensections of the Dirac operator.

 Cutting off the spinor bundle using the Dirac operator of a fixed geometry breaks the symmetry under area preserving diffeomorphisms, but it also introduces power law in distance non-locality into the anti-commutation relations of "local" spinor fields.  We'll argue in the second of these notes that it also preserves a finite dimensional subgroup of the group of area preserving diffeomorphisms.  Such systems certainly do not have a Lieb-Robinson bound, but it is not yet proven that they are fast scramblers.
 
 We want to emphasize one final aspect of this proposal for the dynamical description of causal diamond boundaries before turning to the application of these ideas to cosmology.  It incorporates a long distance short distance duality, reminiscent of Maldacena's scale radius duality in the AdS/CFT correspondence.  The latter applies only to the sphere in the AdS directions, while the present description applies to all transverse dimensions on a given diamond boundary.  The simple fact is that all it is saying is that angular resolution on the sphere becomes better as the radius of the sphere gets larger at fixed UV cutoff.  
 
 However, when combined with the idea that states localized (at some time) near the geodesic in a diamond are constrained states, with the number of constraints a measure of the energy, this principle predicts the qualitative criteria for black hole formation.  If a state is localized near the geodesic at some proper time $t$, then it must be localized in a nested sequence of diamonds.   If it has energy $E$ in a diamond of radius $R$, then it has of order $ER$ constrained q-bits.  But the total number of q-bits is of order $R^{d-2}$, so when $ER \sim R^{d-2}$ this is a highly unlikely state and a small number of actions of the fast scrambling Hamiltonian will replace it with a more generic equilibrium state.   This, with the replacement of fast by ballistic scrambling, is analogous to the fact that a small localized perturbation of an equilibrium state in field theory is relatively long lived, but a state where "all of the gas is clustered in a tiny corner of the room" reverts rapidly to equilibrium.  
 
 In terms of the geometry on the transverse surface, the constrained state with jets of total energy $E$ entering a diamond of radius $R$ is, when $ER \ll R^{d-2}$ a state in which the q-bits in a small area collection of annuli have been set to zero.  The spherical caps inside those annuli are decoupled from the majority of DOF in the system, in that constrained state.  Those DOF are in a highly entangled state, with non-local correlations extending over the system.  As we approach the threshold of classical black hole production, there are roughly the same number of constrained and unconstrained q-bits, and it is no longer possible to use the remnant of area preserving diffeomorphism invariance to separate the constrained annuli from each other.  The action of a term in the Hamiltonian is as likely to turn on a "bridge" between a spherical cap and the unconstrained DOF, as it is to act solely on those unconstrained DOF.  Assuming fast scrambling, in a time of order $R {\rm ln R}$ the system will be in a highly correlated state of all the q-bits, and simple probes of it will behave as they would in a generic equilibrium state.
 The above description is highly qualitative and it is important to demonstrate these statements more explicitly in concrete models.  It is my hope that this note will motivate some younger physicists to do that.
 
 \section{Holographic Inflationary Cosmology}
 
 This section is a brief description of the HST model of inflationary cosmology\cite{holocosm}.  That rather complete reference contains a large number of papers, many of which have errors, which were corrected in later versions of the model.
 The first more or less correct version was\cite{holorevision}.  The basic idea is summarized in the diagram of Figure 2.  This should be thought of as the diagram in conformal coordinates $(\eta, \vec{x})$ of our FRW universe.  
 \begin{figure}[H]
\centering
\includegraphics[scale=0.3]{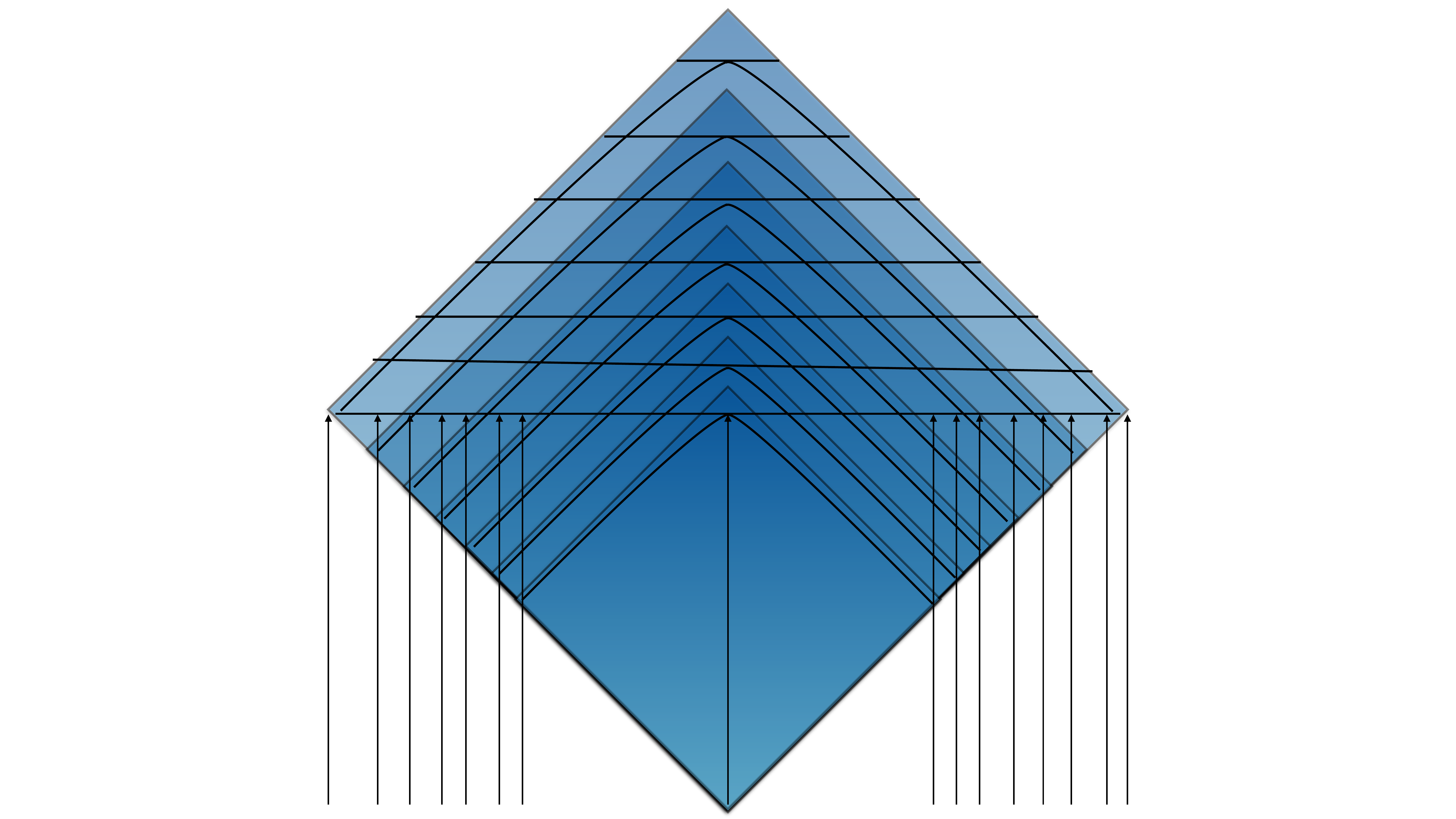}
\caption{Holographic Inflationary Cosmology in Conformal Time}
\label{holoinflation}
\end{figure}
 
 The metric is
 
 \begin{equation} ds^2 = a^2 (\eta) (- d\eta^2 + d\vec{x}^2) .  \end{equation}
 The vertical lines are FRW geodesics, with the one in the center of the diamond being the one followed by our local group of galaxies.  $a(\eta)$ has a pole at the future tip of the diamond, $\eta_0$ representing the presumed asymptotically de Sitter fate of our universe.  $\eta_0/2$ is the end of the slow roll era.   Note that neighboring geodesics enter our past causal boundary at earlier FRW times than the time at which they are observed.  We will argue that the principles of the previous sections imply that if a period of dS expansion was part of the early history of the universe then geodesics that are viewed at a time when their equation of state is $p = - \rho$ appear to us as black holes with average size the inflationary Hubble scale $H_I^{-1}$.  The Hot Big Bang, Baryogenesis, and the CMB spectrum, can all be explained economically by this model.
 It's also possible that these tiny black holes can agglomerate to form primordial black holes (PBHs) with lifetimes that make them candidates for all or part of the Dark Matter.
 
 Let us begin by viewing cosmology from the point of view of a final dS state.   Localized states are constrained and improbable.  For a given total localized energy, the highest probability is to have it all concentrated in a single black hole.  This is indeed likely to be the fate of the universe we observe, if dark energy is really a cosmological constant.  In about $100$ times the current age of the universe, all we will be able to see is the local group of galaxies bound to the Milky Way.  That group, on a much longer time scale, will radiate away its thermal energy and collapse into a black hole.  Finally, that black hole will decay by Hawking radiation.   So the question arises of what the unusual initial condition had to be, which could lead to the diversity of localized objects in the universe we observe.
 
 There is a unique one parameter set of Big Bang FRW cosmologies, which saturates the covariant entropy bound at all times\cite{fsb}.  The spatial universe is flat and the scale factor is
 \begin{equation} a(t) = \sinh^{1/3} (3Ht) , \end{equation} where $H$ is a constant and $t$ is the proper time along a geodesic.  This solves Einstein's equations with a mixture of $p = \rho$ and $p = - \rho$ matter.   We view Einstein's equations as giving us the hydrodynamics of some quantum system\cite{ted95} and search for a quantum system which reproduces the 
hydrodynamic behavior defined by the solution.  Our geometry is singular at $t = 0$, but from the point of view of the CEP this singularity is to be expected.  If the system starts at some initial time, then a few Planck times later its causal diamond has too little entropy for a hydrodynamic description to make sense.  So we just assign Hilbert spaces to later and later instants in time, with an initial random choice of the Hamiltonian which entangles a diamond with the diamond one Planck time later along the same geodesic.  As the diamonds get large, we should match the geometric description, so we postulate\cite{holocosm1+1} that the Hamiltonian approaches that of a $1 +1$ dimensional CFT with UV and IR cutoffs satisfying
\begin{equation} M_c I  = K \gg 1.  \end{equation}   $K$ is time independent, but the central charge of the CFT scales like $t^2$.   The UV cutoff $M_c$ scales like $t^{-1}$, since eigenvalue differences in the Hamiltonian describe the evolution of states on the horizon, as seen from the geodesic.  This is the Milne redshift.   The spatial volume of the interior of the apparent horizon on FRW slices, scales like $t^3$, so the entropy density scales like $t^{-1}$ and the energy density like $t^{-2}$ .  This is the Friedmann equation, with equation of state $p = \rho$!

To incorporate the finite Hubble scale $H$, we simply stop the growth of the Hilbert space at some finite dimension, and let the Hamiltonian approach that of a cutoff $1 + 1$ dimensional CFT with central charge $\sim H^{-2}$.  We can use the formula for area as a function of proper time in the static patch of dS space, to decide how to increase the size of the Hilbert space with each Planck time, but at a certain point this
continuous evolution conflicts with the discreteness of dimension and we just have to let the system evolve in a fixed Hilbert space.  

 In this simple model, we can also solve all the complicated constraints of the QPR.   Choose a different FRW geodesic.  At FRW time $t$ the diamonds along this geodesic and "ours" will have an overlap, containing a maximal diamond.  Our general principles tell us that the modular Hamiltonian of that diamond is the $L_0$ generator of a cutoff $1 + 1$ CFT with central charge determined by the area of the diamond.  We can solve all the consistency conditions by insisting that this CFT be the one which describes the diamond in the FRW past of each trajectory, which has the same area.

It is implicit in this solution that time evolution along any trajectory is described by the {\it same} sequence of time dependent Hamiltonians.  That is, the condition of homogeneity is built in to the definition of our model and has nothing to do with special initial conditions.  Similarly, isotropy is enforced by insisting that every one of our Hamiltonians is invariant under the rotation group, which means in particular that every increase in size of the Hilbert space of a diamond should add a representation of $SU(2)$ to the set of degrees of freedom.   In fact, in our actual models, each Hamiltonian is invariant under the $U(N) \times U(N+1)$ group of unitary transformations on a rectangular matrix of fermionic $\psi_i^A$ oscillators.  It's well known that the action of $U(N)$ on the bilinear $M_i^j = \psi_i^A \psi^{\dagger\ j}_A
$ is a "fuzzy" approximation to the group of area preserving diffeomorphisms of the two sphere.  These groups have many copies of $SU(2)$ as subgroups and we'll see how that gets exploited below.  The increase of $N$ by one unit, scales the entropy exactly in the way we expect the entropy of a diamond to increase in one Planck time.
There are many clues to a general theory of quantum gravity in the combination of this organization of the degrees of freedom, with the idea that the modular Hamiltonian of a diamond is a cutoff version of the $L_0$ generator of a lowest weight representation of the Virasoro algebra, but we don't have time to go into them here.  

It is notable that we did not have to talk about the piece of the time dependent Hamiltonian that acts on degrees of freedom outside a given diamond in order to solve the QPR constraints.  This is a consequence of the fact that our model has no localized excitations, and that the individual Hamiltonians are invariant under (fuzzy) area preserving diffeomorphisms.  There is really no meaning in such a universe to ask "in which direction is that other geodesic with which my diamond has an overlap?". 

  \section{Holographic Inflationary Cosmology}
  
  In the previous section we constructed a model that began with a {\it cold entropic Big Bang}, a $p=\rho$ cosmology, rendered non-singular by quantum mechanics and the CEP, and ended with a dS universe with Hubble constant we will relable $H_I$ .  To construct an inflationary cosmology we must now let the horizon expand, but in such a way that the inflationary Hubble volumes along what were previously causally disconnected geodesics, become independent systems in the expanded causal diamond.   The key picture, in conformal coordinates for our full FRW model, is Figure\ref{holoinflation}.   This picture also incorporates the assumption that the eventual fate of our universe is again a dS space, with Hubble constant $H_{\infty}$.   Our diagram covers only one causal patch of the teleological dS space, with the others viewed as gauge copies\cite{nappietal}\cite{tbwf}.   By contrast, the causally disconnected horizon volumes of our period of inflation, are independent physical systems. A "reasonable" cosmology, which has room for formation of galaxies, requires that $H_{\infty} \ll H_I$.
 A prominent feature of Figure\ref{holoinflation} is that the natural time slices inside "our" causal diamond, intersect geodesics in the FRW past.   The QPR implies that the density matrix of the system that "we" observe entering into our past horizon is the same as the density matrix of dS space.   The modular Hamiltonian is the $L_0$ generator of the same $1 + 1$ dimensional cutoff CFT.  Entropy fluctuations satisfy
 \begin{equation} (\Delta K)^2 = \langle K \rangle . \end{equation}
 
 We now make one further assumption, based the fact that all horizons are locally identical, namely that the inflationary dS CFT is the same one that describes a Schwarzschild black hole with the same horizon area.  Strictly speaking this assumption is not necessary, since we will only use a few properties of black holes in our analysis.  The entropy fluctuations are of course universal without this assumption.  The other thing we will assume is that the localized inflationary horizon volumes move non-relativistically and attract each other with Newtonian forces.   This can probably be derived without making our more microscopic assumption. It's unlikely that experimental physics will ever probe finer details of the quantum mechanics of black holes than those which are guaranteed by general principles of gravitational statistical mechanics.  From now on we will call the isolated horizon volumes (I)nflationary (B)lack (H)oles.  
 
 The requirement that the IBHs remain isolated systems puts two constraints on the period of slow roll, which follows the inflationary era.  Two IBHs will interact to form a single larger black hole if they are too close together, in a time that is 
 \begin{equation} t_{merger} H = - C {\rm ln} (H_I / M_P) , \end{equation}  where the constant $C$ depends on microscopic details.  The slow roll factor must satisfy
 \begin{equation} \epsilon^{-1} \equiv \frac{H^2}{\dot{H}} >  t_{merger} H , \end{equation} in order for the picture of separated IBHs to make sense.  
 
 The bifurcation surface $\eta_0/2$ of the causal diamond of Figure\ref{holoinflation} is the End of Inflation.  It is the maximal spatial slice of the universe that we will ever see, if our cosmology is asymptotically dS with Hubble scale $H_{\infty}$.  In our model, the universe on this slice is a dilute gas of black holes, where the adjective dilute is the second constraint on the slow roll metric that interpolates between this matter dominated universe and inflation.  The separation between black holes must be sufficiently large to avoid constant mergers to produce a maximally entropic $p = \rho$ state.  This is probably at least $3$ Schwarzschild radii, but a more careful computation is necessary.  
 
 I'll now summarize the predictions of this model of inflation and compare them to conventional field theory models.
 \begin{itemize} 
 \item Scalar fluctuations:  the invariant amplitude $\zeta$ is given by
 \begin{equation} \zeta = \epsilon^{-1} \frac{\delta H}{H}  = \epsilon^{-1} (H_I/2 M_P) , \end{equation} where we have used the fluctuation formula for $1 + 1$ CFTs to give an absolute normalization.   This should be compared to 
 \begin{equation} \zeta = \epsilon^{-1/2} (H(t)/M_P) , \end{equation} for a field theory model.   Given our lack of observational probes of the slow roll metric, apart from CMB fluctuations, we can fit either formula to the data.  $\epsilon$ will be larger in the HST fit.  The argument for the approximate scale invariance of the correlations is somewhat different in the HST model than in field theory.  Rotation invariance, and spatial translation invariance in flat FRW coordinates of probability distributions is built into the formalism.   When $(H_I/M_P) $ is small, we also have invariance under scale transformations, that is equivalent to the dS scaling 
 \begin{equation} t \rightarrow t + a, \ \ \ \ \ \vec{x} \rightarrow e^{-a} \vec{x} . \end{equation}  These generate a Borel subalgebra of the $so(1,4)$ Lie algebra.   The remaining generators are obtained by conjugating the spatial translations by the reflection that exchanges the two flat patches that cover global dS space.  This suggests that if we write expectation values in the thermo-field double language, the full dS group is realized.   The Borel subalgebra is sufficient to fix the two point functions to their usual form.  It is notable however that only the $J_{04} $ and $J_{ij}$ generators are realized as Hermitian operators on the quantum Hilbert space of the system.  The invariance under $J_{+i}$ is a consequence of the QPR and is just a constraint on density matrices.  The holographic model also makes different predictions for non-Gaussian fluctuations than field theory models of inflation\cite{holocosmrev}, but because of the small value of $\epsilon$ and Maldacena's squeezed limit theorem\cite{malda}, these fluctuations are not likely to be observed in the near future.  
 
 \item Transplanckian Problems and Swampland Constraints:  the transplanckian problem of field theory inflation models can be restated as the fact that these models postulate that all of the degrees of freedom necessary to explain fluctuations over the entire microwave sky, originated in a single causal diamond of size $H_I^{-1}$.  This violates the CKN\cite{CKN}\cite{bdf}\cite{detal} bound on the field theoretic entropy in a diamond.  The holographic model avoids this problem simply because it insists that modes which are gauge copies of the static patch DOF in a model of dS space, are independent physical systems in a model where a temporary dS phase is followed by a slow roll that expands the size of the horizon.  The Swampland constraints on inflationary models postulate that inflation is caused by a scalar field with a certain type of potential, and that that field can be identified with a modulus of a string theory model in asymptotically flat space.  The holographic models do not make this assumption and Swampland constraints do not apply.  They automatically satisfy unitarity and Einstein locality.  The form of the slow roll metric is holographic models is determined by a choice of the area vs. proper time law in a nested sequence of causal diamonds, the only apparent constraint on slow roll is the bound insuring that IBHs persist as independent systems during slow roll.  Note that this bound on $\epsilon$ would be ruled out if the formula for the power spectrum of fluctuations were proportional to $\epsilon^{-1}$ as it is in field theoretic inflation models.  In HST models the power spectrum is proportional to $\epsilon^{-2}$  and the bound is close to being saturated, within the error bars.  This is probably what we should expect.  Generic initial conditions lead to a universe with no local excitations.  The most probable models with localized excitations are those which just avoid coalescence of those excitations into horizon filling black holes.

\item Tensor fluctuations: the primordial tensor fluctuations have a spectrum that is $SO(1,4)$ invariant up to $(H/M_P)$ corrections.  The tensor to scalar power spectrum ratio is proportional to $\epsilon^2$.  The absolute normalization of the tensor spectrum has not been worked out yet but should follow from the following consideration.  The black hole entropy formula predicts local fluctuations in angular momentum in an inflationary Hubble volume, with a Gaussian width.  This should be compared to the angular momentum fluctuations given by the standard graviton two point function.
There will be another contribution to primordial gravitational waves, which will be produced somewhat later in the history of the universe in the decay of the IBHs, which will set off the Hot Big Bang.  This will be proportional to $1/g$, where $g$ is the number of effectively massless particle species at the decay temperature, and share the spatial profile of the scalar fluctuations.  The decay temperature is calculated by using the black hole lifetime formula and standard cosmology.  It also depends on the assumed energy density at the end of inflation.  This is $\rho_I = K m^{-2}$, where $K$ is determined by the requirement that the black hole number density is just\footnote{The word {\it just} is motivated by the requirement that our initial conditions be the most probable for which the dilute gas picture and Hot Big Bang occur.} low enough to avoid merging into a $p = \rho$ phase.  The average IBH mass $m = R_I/2$ in Planck units and the observed normalization of the scalar fluctuations gives 
\begin{equation} 10^{-5} = (\sqrt{\pi} R_I \epsilon)^{-1}  . \end{equation}  In making quantitative estimates we have generally taken $\epsilon = 0.1$, which gives $m \sim 10^6$.   This gives a black hole evaporation temperature of order $10^{10} - 10^8$ GeV, taking into account the various uncertainties.

\item Baryogenesis:  Black hole formation and evaporation does not conserve baryon or lepton number, and the change in black hole mass $dM/dt$ violates CPT.  Assuming CP violation of order $1$ in the evaporation process one gets a crude estimate\cite{tbwfbaryo} within the right ballpark for the baryon to entropy ratio at the Hot Big Bang.  

\item  The IBH decay time is of order $t_{decay} \sim m^3$, while the size of primordial fluctuations is $ \sim (m\epsilon)^{-1} $.  During the dilute black hole gas phase of the universe, these fluctuations grow like $t^{2/3}$ , so become of order $1$ at a time $t_{merge} \sim (m \epsilon)^{3/2}$, parametrically earlier than the decay.  This means that it's highly likely that some longer lived black holes will be formed by mergers.  During a radiation dominated phase of the universe, these black holes will grow by an order $1$ factor until the radiation temperature falls below their Hawking temperature\cite{carrhawking} and then decay.  A black hole of mass $10^{11}$ or greater will survive the radiation dominated era.  It turns out that we need a probability of only $10^{-24}$ that $10^{5}$ IBHs will merge to form one of these P(rimordial) B(lack) H(ole)s, in order to account for all of the dark matter in the universe at the observed moment of matter domination.  From that point on, the scenario becomes more complex and computer simulations are necessary to determine whether these PBHs can combine quickly enough to form cosmologically stable structures before they decay.   Decay of more than a few percent of them would be inconsistent with observation.  
\item  A recent paper\cite{???} has claimed that if each IBH leaves over a Planck mass remnant, then models of this type might account for dark matter without the worries of merger histories.  I have not studied this proposal with sufficient care, but it is an interesting new direction in holographic cosmology.
\end{itemize}

In summary, a very economical set of models, based on the principles of the previous section, can plausibly explain everything we know observationally about the very early universe, with the possible exception of the origin of dark matter.  All of the calculations sketched here need to be refined.  Most pressing and most difficult is the question of whether these models can generate a consistent theory of PBH dark matter.

\section{Conclusions}

Analysis of semi-classical  entropy formulae, and quasi-normal modes indicates a universal local behavior of null surfaces in models of quantum gravity.   Semi-classical arguments\cite{CKN} and QFT calculations of the entanglement entropy of causal diamonds\cite{sred}\cite{cw} suggest that the Bekenstein-Hawking formula for black hole entropy applies to the entropy of any causal diamond, identifying a finite area diamond as a finite dimensional subsystem in any model of QG.  The work of\cite{carlip} and \cite{solodukhin} suggests a universal formula for entropy fluctuations in any diamond\cite{BZ}.

The causal structure of QG is then identical to that of QFT, with only the relation between the total entropy and geometry distinguishing them.  This leads to a formulation of QG in terms of "many-fingered" proper time evolution along different time-like trajectories.  Causality is implemented by proper time dependent Hamiltonians, which couple together only the DOF in the causal diamond of a given time interval.   The principle of relativity is implemented by the requirement that the density matrices assigned to the maximal diamond in the intersection of diamonds along two different trajectories, have identical entanglement spectra.

The above principles localize quantum information on the boundaries of causal diamonds, with Hamiltonian evolution entangling the q-bits in one diamond with the new ones that are added every Planck time.  They do not tell us how to describe excitations localized in the bulk of a diamond.  Black hole entropy formulae suggest that such localized states are constrained states of the boundary system, with a number of constrained q-bits scaling like $ER$ where $R$ is the radius of the diamond and $E$ the localized energy.  The combination of these principles leads to a holographic formulation of inflationary cosmology, which does not suffer from "trans-Planckian" problems and gives rise to a post-inflationary dilute gas of black holes, whose average size is the inflationary Hubble scale.  This model can fit all extant data on the cosmology of the very early universe, and might even lead to a consistent model of primordial black hole dark matter.

This note has summarized years of work on a general theory of quantum gravity that explicitly incorporates locality into its axioms.  That theory, in its present form, is cumbersome and does not lead to elegant methods of calculation.  It is the author's hope that some younger researchers will be motivated to improve it.   I have tried to rely only on results and conjectures that are clearly based on semi-classical reasoning and indicate the way in which these conjectures are consistent with what we actually know about the utility of QFT for describing the real world.   The most important result, so far, of this formalism is a completely well defined set of models for inflationary cosmology, which are very economical, consistent with data, and falsifiable.  Their predictions differ from those of QFT models of inflation, but not at a level that is currently testable.  The models might fail to explain Dark Matter.

In a companion note I will indicate how the same set of ideas, with a bit of extra input, lead to insights about particle physics Beyond the Standard Model.


\begin{thebibliography}{99}
\bibitem{tbwf} W.~Fischler, {\it Taking de Sitter Seriously}, Talk at the Festschrift for Geoffrey West, Taos, New Mexico,    2000.
T.~Banks, Talk at the Festschrift for L. Susskind, Stanford University, May 2000.   T.~Banks, ``Cosmological breaking of supersymmetry?,''
Int. J. Mod. Phys. A \textbf{16}, 910-921 (2001)
doi:10.1142/S0217751X01003998
[arXiv:hep-th/0007146 [hep-th]].

\bibitem{bfm} T.~Banks, B.~Fiol and A.~Morisse, ``Towards a quantum theory of de Sitter space,''
JHEP \textbf{12}, 004 (2006)
doi:10.1088/1126-6708/2006/12/004
[arXiv:hep-th/0609062 [hep-th]].
\bibitem{HST} 
T.~Banks and W.~Fischler, ``Holographic Inflation Revised,''
doi:10.1017/9781316535783.013
[arXiv:1501.01686 [hep-th]].
T.~Banks and W.~Fischler, ``Holographic Space-time, Newton's Law and the Dynamics of Black Holes,''
[arXiv:1606.01267 [hep-th]].
T.~Banks and W.~Fischler, ``Holographic Space-time Models of Anti-deSitter Space-times,''
[arXiv:1607.03510 [hep-th]].
T.~Banks and W.~Fischler, ``The holographic spacetime model of cosmology,''
Int. J. Mod. Phys. D \textbf{27}, no.14, 1846005 (2018)
doi:10.1142/S0218271818460057
[arXiv:1806.01749 [hep-th]].
T.~Banks and W.~Fischler, ``Entropy and Black Holes in the Very Early Universe,''
[arXiv:2109.05571 [hep-th]], and references in all of the above.
\bibitem{ETH} J. M. Deutsch, Phys. Rev. A 43, 2046 (1991).
M. Srednicki, Phys. Rev. E 50, 888 (1994).
\bibitem{CKN} A.~G.~Cohen, D.~B.~Kaplan and A.~E.~Nelson,
``Effective field theory, black holes, and the cosmological constant,''
Phys. Rev. Lett. \textbf{82}, 4971-4974 (1999)
doi:10.1103/PhysRevLett.82.4971
[arXiv:hep-th/9803132 [hep-th]].
\bibitem{sred}M.~Srednicki, ``Entropy and area,''
Phys. Rev. Lett. \textbf{71}, 666-669 (1993)
doi:10.1103/PhysRevLett.71.666
[arXiv:hep-th/9303048 [hep-th]].
\bibitem{cw}C.~G.~Callan, Jr. and F.~Wilczek, ``On geometric entropy,''
Phys. Lett. B \textbf{333}, 55-61 (1994)
doi:10.1016/0370-2693(94)91007-3
[arXiv:hep-th/9401072 [hep-th]].
\bibitem{sorkin} R. Sorkin, ``On The Entropy of the Vacuum Outside a Horizon," in B. Bertotti, F. de Felice and A. Pascolini, eds., Tenth International Conference on General Relativity and Gravitation (Padova, July 4-9, 1983), Contributed Papers, vol. II, pp. 734-736 (Roma, Consiglio Nazionale Delle Ricerche, 1983), available at arXiv:1402.3589.
\bibitem{CHM} H.~Casini, M.~Huerta and R.~C.~Myers, ``Towards a derivation of holographic entanglement entropy,''
JHEP \textbf{05}, 036 (2011)
doi:10.1007/JHEP05(2011)036
[arXiv:1102.0440 [hep-th]].
\bibitem{JV} T.~Jacobson and M.~Visser, ``Gravitational Thermodynamics of Causal Diamonds in (A)dS,''
SciPost Phys. \textbf{7}, no.6, 079 (2019)
doi:10.21468/SciPostPhys.7.6.079
[arXiv:1812.01596 [hep-th]].

\bibitem{liebrobinson} E. H. Lieb and D. W. Robinson
Commun. Math. Phys. 28, 251 (1972).
\bibitem{tn} B.~Swingle,
``Entanglement Renormalization and Holography,''
Phys. Rev. D \textbf{86}, 065007 (2012)
doi:10.1103/PhysRevD.86.065007
[arXiv:0905.1317 [cond-mat.str-el]].


\bibitem{su} L.~Susskind, L.~Thorlacius and J.~Uglum,
``The Stretched horizon and black hole complementarity,''
Phys. Rev. D \textbf{48}, 3743-3761 (1993)
doi:10.1103/PhysRevD.48.3743
[arXiv:hep-th/9306069 [hep-th]].
\bibitem{ct} S.~Carlip and C.~Teitelboim,``The Off-shell black hole,''
Class. Quant. Grav. \textbf{12}, 1699-1704 (1995)
doi:10.1088/0264-9381/12/7/011
[arXiv:gr-qc/9312002 [gr-qc]].
\bibitem{bdf} T.~Banks, P.~Draper and S.~Farkas,
``Path Integrals for Causal Diamonds and the Covariant Entropy Principle,''
Phys. Rev. D \textbf{103}, no.10, 106022 (2021)
doi:10.1103/PhysRevD.103.106022
[arXiv:2008.03449 [hep-th]].
\bibitem{bd}T.~Banks and P.~Draper,
``Remarks on the Cohen-Kaplan-Nelson bound,''
Phys. Rev. D \textbf{101}, no.12, 126010 (2020)
doi:10.1103/PhysRevD.101.126010
[arXiv:1911.05778 [hep-th]].
\bibitem{detal} N.~Blinov and P.~Draper, ``Densities of states and the Cohen-Kaplan-Nelson bound,''
Phys. Rev. D \textbf{104}, no.7, 076024 (2021)
doi:10.1103/PhysRevD.104.076024
[arXiv:2107.03530 [hep-ph]].

\bibitem{fsb} W.~Fischler and L.~Susskind, ``Holography and cosmology,''
[arXiv:hep-th/9806039 [hep-th]].
R.~Bousso, ``A Covariant entropy conjecture,''
JHEP \textbf{07}, 004 (1999)
doi:10.1088/1126-6708/1999/07/004
[arXiv:hep-th/9905177 [hep-th]].

\bibitem{ted95} T.~Jacobson, ``Thermodynamics of space-time: The Einstein equation of state,''
Phys. Rev. Lett. \textbf{75}, 1260-1263 (1995)
doi:10.1103/PhysRevLett.75.1260
[arXiv:gr-qc/9504004 [gr-qc]].


\bibitem{gh1}G.~W.~Gibbons and S.~W.~Hawking,  ``Action Integrals and Partition Functions in Quantum Gravity,''
Phys. Rev. D \textbf{15}, 2752-2756 (1977)
doi:10.1103/PhysRevD.15.2752

\bibitem{gh2}G.~W.~Gibbons and S.~W.~Hawking,
``Cosmological Event Horizons, Thermodynamics, and Particle Creation,''
Phys. Rev. D \textbf{15}, 2738-2751 (1977)
doi:10.1103/PhysRevD.15.2738
\bibitem{cdl} S.~R.~Coleman and F.~De Luccia,``Gravitational Effects on and of Vacuum Decay,''
Phys. Rev. D \textbf{21}, 3305 (1980)
doi:10.1103/PhysRevD.21.3305
\bibitem{tbheretic} T.~Banks, ``Heretics of the false vacuum: Gravitational effects on and of vacuum decay. 2.,''
[arXiv:hep-th/0211160 [hep-th]].
\bibitem{bw}A.~R.~Brown and E.~J.~Weinberg, ``Thermal derivation of the Coleman-De Luccia tunneling prescription,''
Phys. Rev. D \textbf{76}, 064003 (2007)
doi:10.1103/PhysRevD.76.064003
[arXiv:0706.1573 [hep-th]].
\bibitem{df} P.~Draper and S.~Farkas,
``Euclidean de Sitter black holes and microcanonical equilibrium,''
Phys. Rev. D \textbf{105}, no.12, 126021 (2022)
doi:10.1103/PhysRevD.105.126021
[arXiv:2203.01871 [hep-th]].
 P.~Draper and S.~Farkas,
``de Sitter black holes as constrained states in the Euclidean path integral,''
Phys. Rev. D \textbf{105}, no.12, 126022 (2022)
doi:10.1103/PhysRevD.105.126022
[arXiv:2203.02426 [hep-th]].
\bibitem{maldleuk}A.~Lewkowycz and J.~Maldacena, ``Generalized gravitational entropy,''
JHEP \textbf{08}, 090 (2013)
doi:10.1007/JHEP08(2013)090
[arXiv:1304.4926 [hep-th]].
\bibitem{bl} T.~Banks and A.~Lucas, ``Emergent entropy production and hydrodynamics in quantum many-body systems,''
Phys. Rev. E \textbf{99}, no.2, 022105 (2019)
doi:10.1103/PhysRevE.99.022105
[arXiv:1810.11024 [cond-mat.stat-mech]].
\bibitem{hong} H.~Liu and P.~Glorioso, ``Lectures on non-equilibrium effective field theories and fluctuating hydrodynamics,''
PoS \textbf{TASI2017}, 008 (2018)
doi:10.22323/1.305.0008
[arXiv:1805.09331 [hep-th]].
\bibitem{ranga} F.~M.~Haehl, R.~Loganayagam and M.~Rangamani,
``The Fluid Manifesto: Emergent symmetries, hydrodynamics, and black holes,''
JHEP \textbf{01}, 184 (2016)
doi:10.1007/JHEP01(2016)184
[arXiv:1510.02494 [hep-th]].
\bibitem{page}D.~N.~Page, ``Average entropy of a subsystem,''
Phys. Rev. Lett. \textbf{71}, 1291-1294 (1993)
doi:10.1103/PhysRevLett.71.1291
[arXiv:gr-qc/9305007 [gr-qc]].
\bibitem{ecc} A.~Almheiri, X.~Dong and D.~Harlow, ``Bulk Locality and Quantum Error Correction in AdS/CFT,''
JHEP \textbf{04}, 163 (2015)
doi:10.1007/JHEP04(2015)163
[arXiv:1411.7041 [hep-th]].

\bibitem{tbwfhigherd} T.~Banks and W.~Fischler, ``Holographic Space-time, Newton's Law and the Dynamics of Black Holes,''
[arXiv:1606.01267 [hep-th]].

\bibitem{gkpw} S.~S.~Gubser, I.~R.~Klebanov and A.~M.~Polyakov,
``Gauge theory correlators from noncritical string theory,''
Phys. Lett. B \textbf{428}, 105-114 (1998)
doi:10.1016/S0370-2693(98)00377-3
[arXiv:hep-th/9802109 [hep-th]].
E.~Witten, ``Anti-de Sitter space and holography,''
Adv. Theor. Math. Phys. \textbf{2}, 253-291 (1998)
doi:10.4310/ATMP.1998.v2.n2.a2
[arXiv:hep-th/9802150 [hep-th]].
\bibitem{witsuss} L.~Susskind and E.~Witten,
``The Holographic bound in anti-de Sitter space,''
[arXiv:hep-th/9805114 [hep-th]].


\bibitem{scaleradius}J.~M.~Maldacena,
``The Large N limit of superconformal field theories and supergravity,''
Adv. Theor. Math. Phys. \textbf{2}, 231-252 (1998)
doi:10.1023/A:1026654312961
[arXiv:hep-th/9711200 [hep-th]].
\bibitem{tn}B.~Swingle,
``Entanglement Renormalization and Holography,''
Phys. Rev. D \textbf{86}, 065007 (2012)
doi:10.1103/PhysRevD.86.065007
[arXiv:0905.1317 [cond-mat.str-el]].


\bibitem{tnrg}R.~N.~C.~Pfeifer, G.~Evenbly and G.~Vidal,
``Entanglement renormalization, scale invariance, and quantum criticality,''
Phys. Rev. A \textbf{79}, 040301 (2009)
doi:10.1103/PhysRevA.79.040301
[arXiv:0810.0580 [cond-mat.str-el]].
\bibitem{tbwfads}T.~Banks and W.~Fischler, ``Holographic Space-time Models of Anti-deSitter Space-times,''
[arXiv:1607.03510 [hep-th]].
\bibitem{malda} J.~M.~Maldacena,``Non-Gaussian features of primordial fluctuations in single field inflationary models,''
JHEP \textbf{05}, 013 (2003)
doi:10.1088/1126-6708/2003/05/013
[arXiv:astro-ph/0210603 [astro-ph]].

\bibitem{susslargenode}    L.~Susskind, {\it Entanglement is Not Enough},
Fortsch.Phys. 64 (2016) 49-71,e-Print: 1411.0690 [hep-th].
\bibitem{BZ} T.~Banks and K.~M.~Zurek, ``Conformal description of near-horizon vacuum states,''
Phys. Rev. D \textbf{104}, no.12, 126026 (2021)
doi:10.1103/PhysRevD.104.126026
[arXiv:2108.04806 [hep-th]].
\bibitem{carlip} S.~Carlip, ``Statistical mechanics and black hole entropy,''
[arXiv:gr-qc/9509024 [gr-qc]];
Phys. Rev. D 51, 632 (1995), arXiv:gr-qc/9409052;
Phys. Rev. Lett. 82, 2828 (1999), arXiv:hep-th/9812013;
Class. Quant. Grav. 15, 3609 (1998), arXiv:hep-th/9806026;
AIP Conf. Proc. 1483, 54 (2012), arXiv:1207.1488 [gr-qc].

\bibitem{solodukhin}S. N. Solodukhin, Phys. Lett. B 454, 213 (1999), arXiv:hep-th/9812056.
\bibitem{andy} A. Strominger, JHEP 02, 009 (1998), arXiv:hep-th/9712251.
\bibitem{brownhenneaux} J. D. Brown and M. Henneaux, Commun. Math. Phys. 104, 207 (1986). 
\bibitem{VZ2} E. Verlinde and K. M. Zurek, JHEP 04, 209 (2020), arXiv:1911.02018 [hep-th].
\bibitem{RS2} L. Randall and R. Sundrum, Phys. Rev. Lett. 83, 4690 (1999), arXiv:hep-th/9906064.
\bibitem{perlprivate} E. Perlmutter, JHEP 03, 117 (2014), arXiv:1308.1083 [hep-th], and private communication to K. Zurek.
\bibitem{membrane} {\it Black Holes: the Membrane Paradigm}, eds. D.H. MacDonald, R.H. Price, K.S. Thorne,  ISBN-13: 978-0300037708
ISBN-10: 0300037708  Yale University Press, 1986.
\bibitem{susslind} L. Susskind, J. Lindesay, {\it Introduction to Black Holes, Information, and the String Theory Revolution}, World Scientific, 2005.

\bibitem{hp} P. Hayden and J. Preskill, JHEP 09, 120 (2007), arXiv:0708.4025 [hep-th].

\bibitem{ss} Y. Sekino and L. Susskind, JHEP 10, 065 (2008), arXiv:0808.2096 [hep-th].
\bibitem{bfss} T.~Banks, W.~Fischler, S.~H.~Shenker and L.~Susskind, ``M theory as a matrix model: A Conjecture,''
Phys. Rev. D \textbf{55}, 5112-5128 (1997)
doi:10.1103/PhysRevD.55.5112
[arXiv:hep-th/9610043 [hep-th]].
\bibitem{holocosm} T.~Banks and W.~Fischler, ``An Holographic cosmology,''
[arXiv:hep-th/0111142 [hep-th]]; 
T.~Banks and W.~Fischler, ``Holographic cosmology 3.0,''
Phys. Scripta T \textbf{117}, 56-63 (2005)
doi:10.1238/Physica.Topical.117a00056
[arXiv:hep-th/0310288 [hep-th]];
\bibitem{Banks:2004vg}
T.~Banks and W.~Fischler, ``Holographic cosmology,''
[arXiv:hep-th/0405200 [hep-th]];
T.~Banks, W.~Fischler and L.~Mannelli, ``Microscopic quantum mechanics of the p = rho universe,''
Phys. Rev. D \textbf{71}, 123514 (2005)
doi:10.1103/PhysRevD.71.123514
[arXiv:hep-th/0408076 [hep-th]];
T.~Banks and W.~Fischler, ``The holographic approach to cosmology,''
[arXiv:hep-th/0412097 [hep-th]];
T.~Banks and W.~Fischler, ``Holographic Theories of Inflation and Fluctuations,''
[arXiv:1111.4948 [hep-th]];
T.~Banks, W.~Fischler, T.~J.~Torres and C.~L.~Wainwright,
``Holographic Fluctuations from Unitary de Sitter Invariant Field Theory,''
[arXiv:1306.3999 [hep-th]];
T.~Banks and W.~Fischler, ``Holographic Inflation Revised,''
doi:10.1017/9781316535783.013
[arXiv:1501.01686 [hep-th]];
T.~Banks and W.~Fischler, ``CP Violation and Baryogenesis in the Presence of Black Holes,''
[arXiv:1505.00472 [hep-th]];
T.~Banks and W.~Fischler, ``The holographic spacetime model of cosmology,''
Int. J. Mod. Phys. D \textbf{27}, no.14, 1846005 (2018)
doi:10.1142/S0218271818460057
[arXiv:1806.01749 [hep-th]];
\bibitem{Banks:2020dgx}
T.~Banks and W.~Fischler, ``Primordial Black Holes as Dark Matter,''
[arXiv:2008.00327 [hep-th]];
T.~Banks and W.~Fischler, ``Entropy and Black Holes in the Very Early Universe,''
[arXiv:2109.05571 [hep-th]].

\bibitem{holorevision} T.~Banks and W.~Fischler, ``Holographic Inflation Revised,''
doi:10.1017/9781316535783.013
[arXiv:1501.01686 [hep-th]];
\bibitem{nappietal} R.~Figari, R.~Hoegh-Krohn and C.~R.~Nappi,
``Interacting Relativistic Boson Fields in the de Sitter Universe with Two Space-Time Dimensions,''
Commun. Math. Phys. \textbf{44}, 265-278 (1975)
doi:10.1007/BF01609830

\bibitem{holocosm1+1} T.~Banks, W.~Fischler and L.~Mannelli, ``Microscopic quantum mechanics of the p = rho universe,''
Phys. Rev. D \textbf{71}, 123514 (2005)
doi:10.1103/PhysRevD.71.123514
[arXiv:hep-th/0408076 [hep-th]];
\bibitem{tbwfbaryo}T.~Banks and W.~Fischler, ``CP Violation and Baryogenesis in the Presence of Black Holes,''
[arXiv:1505.00472 [hep-th]].

\bibitem{carrhawking} B.~J.~Carr and S.~W.~Hawking, ``Black holes in the early Universe,''
Mon. Not. Roy. Astron. Soc. \textbf{168}, 399-415 (1974)
T.~Banks and W.~Fischler, ``Primordial Black Holes as Dark Matter,''
[arXiv:2008.00327 [hep-th]];
T.~Banks and W.~Fischler, ``Entropy and Black Holes in the Very Early Universe,''
[arXiv:2109.05571 [hep-th]].
\bibitem{???}A.~Barrau, ``The holographic space-time and black hole remnants as dark matter,''
Phys. Lett. B \textbf{829}, 137061 (2022)
doi:10.1016/j.physletb.2022.137061
[arXiv:2201.06988 [gr-qc]].
\bibitem{V291} H.~L.~Verlinde and E.~P.~Verlinde, ``Scattering at Planckian energies,''
Nucl. Phys. B \textbf{371}, 246-268 (1992)
doi:10.1016/0550-3213(92)90236-5
[arXiv:hep-th/9110017 [hep-th]].


\end{thebibliography}
\end{document}